# Influence of solvent polarization and non-uniform ionic size on electrokinetic transport in a nanochannel

Jun-Sik Sin[1, 2,*], Nam-Hyok Kim[1], Chol-Ho Kim[2], Yong-Man Jang[2]

[1]Department of Physics, Kim Il Sung University, Pyongyang, Democratic People's Republic of Korea
[2]Natural Science Center, Kim Il Sung University, Pyongyang, Democratic People's Republic of Korea

* js.sin@ryongnamsan.edu.kp

Abstract: In this paper, we study the electroosmotic transport in a nanofluidic channel by using a mean-field theory accounting for non-uniform size effect and solvent polarization effect. We witness that in the presence of the given zeta potential, an enhancement of ion size invariably lowers the electroosmotic velocity, thereby increasing the magnitude of electrostatic potential, irrespective of considering solvent polarization. It is also proved that solvent polarization allows both the magnitude of electrostatic potential and the electroosmotic velocities to decrease. In addition, we find that increasing zeta potential augments not only ion size effect but also solvent polarization effect. Furthermore, we demonstrate that decreasing bulk ion number density causes an increase in electroosmotic velocity at the centerline. We compare the properties of aqueous electrolytes with those of the electrolytes where solvent is ethylalcohol. Finally, we study how solvent polarization and ionic size affect streaming potential and electroviscous effect. It is emphasized that the present study can provide a good way to control the nanofluidic transport for a plethora of biological and industrial applications.

1. Introduction

For the last decades, vast studies have been devoted to studying electroosmotic flow (Hunter 1981; Tabeling 2005; Ghosal 2006) in nanofluidic channels since electroosmotic flow can convey liquid and liquid-driven moieties at much less energy than that for the traditional pressure-driven transport.

In order to successfully realize a number of applications related to the electric field driven transport in nanochannels, it is necessary for us to screen/suppress electroosmotic transport that is triggered by the interaction of the electric double layer and the axial applied electric field.

Therefore, understanding electroosmotic transport in nanofluidic channels requires the development of a reasonable model of electric double layer that is the nanometer-scale layer of electric charges located at the interface between a charged object and an electrolyte.

The classic electric double layer model is Gouy-Chapman model based on Poisson-Boltzmann approach (PB), which disregards finite ion size, ion-ion correlation and solvent polarization. Poisson-Boltzmann approach was used to investigate the streaming potential (Chakraborty and Das 2008; Das and Chakraborty 2009, 2010; Das et al. 2013, Chanda et al. 2014; Chen and Das 2015) and electroviscous effects (Chakraborty and Das 2008; Das et al. 2013; Chanda et al. 2014; Chen and Das 2015) in nanochannels.

In order to overcome the drawbacks of PB approach, many studies have been attempted to develop electric double layer approaches including different effects such as finite ionic size effects (Kilic and Bazant 2007; Bazant et al. 2009; Lamperski and Outhwaite 2002; Das and Chakraborty 2011; Das et al. 2012a, b; Chanda and Das 2014; Bohinc et al. 2001; Borukhov et al. 1997; Boschitsch and Danilov 2012; Popovic and Siber 2013) , ion - ion and ion-solvent interactions (Storey and Bazant 2012; Levine and Feat 1977; Kjellander 2009; Bazant et al. 2011; Bohinc et al. 2012; Zhao 2012), and solvent polarization effects (Iglic et al. 2010; Gongadze and Iglic 2012; Outhwaite 1976, 1983; Misra et al 2013,) on electric double layer electrostatics.

Such a development of electric double layer model provoked improvement in modeling electrokinetic transport in nanochannels, considering the above-mentioned effects. First and foremost, it should be mentioned that (Storey et al 2008; Bandopadhyay and Chakraborty 2013) studied the effect of finite ion sizes in electroosmotic transport.

The authors of (Bandopadhyay et al. 2013, 2014; Dhar et al. 2014; Bandopadhyay and Dhar 2013; Bandopadhyay and Chakraborty 2015) also investigated influence of ionic sizes and non-electrostatic ion-ion correlations on the streaming potential and electroviscous effects in nanochannels. Although ionic size effect and ion correlation effect on electrokinetic transport have been considered, all the studies didn't address solvent polarization effect on electrokinetic phenomena.

In a recent study (Sinha et al. 2015), Das and co-workers studied the role of solvent polarization in determining electroosmotic transport. However, their model that assumes an identical size of ions and solvent molecules is inapplicable to realistic cases where ions and solvent molecules have different sizes.

On the other hand, the authors of (Sin et al. 2015; Gongadze and Iglic 2015; Sin et al. 2016a) presented mean-field theories accounting for solvent polarization effect and non-uniform size effect on electric double layer near a charged surface in an electrolyte.

Based on these approaches, we studied several interesting phenomena related to multicomponent electrolytes (Sin et al. 2016a), binary polar solvent mixtures (Sin et al. 2016b), soft charged surfaces (Sin et al. 2017) and electrostatic interaction between charged surfaces (Sin 2017).

In this paper, we study the electroosmotic transport in a nanochannel for the case where the solvent is polarized by the electric double layer electric field and a solvent molecule has the different size from that for ions. The electric double layer model is based on a mean-field approach (Sin et al. 2016) taking ionic size and orientational ordering of solvent molecules into account simultaneously. When solvent polarization is disregarded, our model is reduced to the different mean-field approaches considering only the ion size effect (Bohinc et al. 2001; Borukhov et el. 1997; Boschitsch and Danilov 2012; Popovic and Siber 2013). This model is also a generalization of Langevin-Bikermann(LB) model (Iglic et al. 2010) that considered the solvent polarization and an equal size of ions and solvent molecules.

It is worthwhile to mention that the solvent polarization effect has been extensively used to better represent biological interactions (Luque et al. 1995), quantify electric double layer - mediated disjoining pressure between nanoscopically separated charged surfaces (Misra et al 2013; Sin 2017), study nitroethane deprotonation (Major et al. 2005).

In fact, in our previous paper (Sin 2017), we anticipated that the present model provides a realistic representation of electric double layer interaction force between similar or oppositely charged surfaces separated by nanometer gap.

In this paper, we shall probe how simultaneous consideration of solvent polarization effect and ion size effect affects electrokinetic transport in a nanochannel. We also study roles of zeta potential and bulk ion concentration in determining electrostatic and electroosmotic properties. More importantly, we establish the relationship between the effects of ion size, solvent polarization and zeta potential on electrostatic and electroosmotic properties. A comparison with the molecular dynamics result (Kim and Darve 2006) verifies the validity of the present mean-field model. In addition, we study the influence of solvent polarization and ionic size on streaming potential and electroviscous effect.

**2. Theory**

2.1 Electroosmotic transport

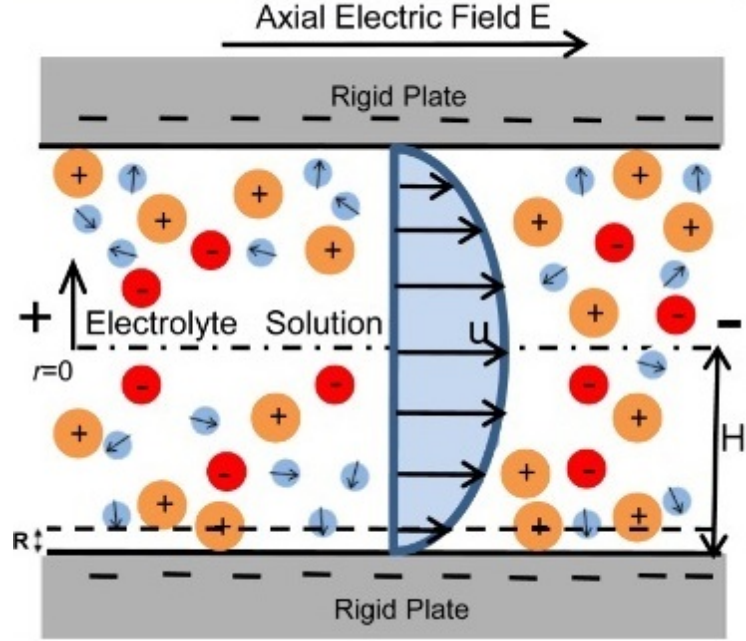

Fig. 1. (Color online) Schematic diagram of the electroosmotic transport in a nanochannel. The electrolyte cations and anions are shown in yellow and red, respectively, while the water dipoles are shown in blue.

As shown in Fig. (1), we consider an axial electric field driven electroosmotic flow in a nanofluidic channel. The height of the nanochannel is $2H$. Our model accounts for the effect of solvent polarization and non-uniform ion size on the electroosmotic transport.

The transverse direction is noticed by $r$; the bottom plate is placed at $r=-H$ and the top plate $r=H$.

The electrostatic properties at the interface between the plates and the electrolyte solution should be addressed by setting the free energy of total thermodynamic system as follows:

$$F = \int d\mathbf{r} \left[ -\frac{\varepsilon_0 (\nabla \psi)^2}{2} + e_0 \psi(\mathbf{r})(n_+ - n_-) + \langle p_0 |\nabla \psi| \cos \omega \rho(\omega) \rangle_\omega - \mu_+ n_+ - \mu_- n_- - \langle \mu_\omega(\omega) \rho(\omega) \rangle_\omega - Ts \right]$$

(1)

While the local electrostatic potential is denoted by $\psi$, the number density of different ionic species and the number density of water molecules are expressed as $n_i(r)$, (i=+, -) and $n_w(\mathbf{r}) = \langle \rho(\omega, \mathbf{r}) \rangle_\omega$, respectively.

Here $\langle f(\omega) \rangle_\omega = \int f(\omega) 2\pi \sin(\omega) d\omega$. $\omega$ stands for the angle between the dipole moment of a water molecule p and the normal to the charged surface. $P_0 = |p|$.

In Eq. (1), the first term describes the self energy of the electrostatic field, where $\varepsilon_r$ stands for the vacuum permittivity. The second term means the electrostatic energy of the ions. It is noticeable that the electrostatic energy of water dipoles being the third term of Eq. (1) is equal to one of (Iglic et al. 2010).

The next three terms are needed to couple the system to a bulk reservoir, where $\mu_i$ means the chemical potential of the ions and $\mu_w(\omega)$ corresponds to the chemical potential of water dipoles with orientational angle $\omega$. $T$ and $s$ are the temperature and the entropy density, respectively.

From this free energy functional, we derive the self-consistent equations determining electrostatic properties by performing minimization of the corresponding free energy describing the electric double layer and consequently find the formula of the osmotic pressure.

The Lagrangian of the electrolyte with an undetermined multiplier can be expressed using the volume conservation as constraint condition

$$L = F - \int \lambda(\mathbf{r})(1 - n_+ V_+ - n_- V_- - n_w V_w) d\mathbf{r} , \qquad (2)$$

where $\lambda$ is a local Lagrange parameter.

The number densities of ions and water molecules can be obtained by combining Eqs. (1) and (2):

$$n_+ = \frac{n_{+b} \exp(-V_+ h - e_0 \beta \psi)}{D}$$

$$n_- = \frac{n_{-b} \exp(-V_- h + e_0 \beta \psi)}{D}$$

$$n_w = \frac{n_{wb} \exp(-V_w h) \dfrac{\sinh(p_0 \beta |\nabla \psi|)}{p_0 \beta |\nabla \psi|}}{D} , \qquad (3)$$

where $D = n_{+b} V_+ \exp(-V_+ h - e_0 \beta \psi) + n_{-b} V_- \exp(-V_- h + e_0 \beta \psi) + n_{wb} V_w \dfrac{\sinh(p_0 \beta |\nabla \psi|)}{p_0 \beta |\nabla \psi|}$.

(please refer to (Sin 2017))

In addition to the above equations, another equation is derived by combining Eqs. (1) and (2):

$$n_{+b}\left(e^{-V_+h-e_0\beta\phi}-1\right)+n_{-b}\left(e^{-V_-h+e_0\beta\phi}-1\right)+n_{wb}\left(e^{-V_wh}\frac{\sinh(p_0\beta|\nabla\psi|)}{p_0\beta|\nabla\psi|}-1\right)=0. \quad (4)$$

Euler-Lagrange equation for $\psi$ yields the Poisson equation:

$$\nabla(\varepsilon_0\varepsilon_r\nabla\psi)=-e_0(n_+-n_-), \qquad r>0 \quad (5)$$

where $\varepsilon_r \equiv 1+\dfrac{|\mathbf{P}|}{\varepsilon_0|\nabla\psi|}$.

Due to planar symmetry of the present study, P, the polarization vector due to a total orientation of point-like water dipoles, is perpendicular to charged surfaces, as in

(Iglic et al. 2010; Gongadze and Iglic 2012; Sin et al. 2015; Sin et al. 2016)

$$\mathbf{P}(r)=n_w(r)p_0 L(p_0|\nabla\psi|\beta), \quad (6)$$

where $\beta=1/(k_BT)$ and a function $L(u)=\coth(u)-1/u$ means the Langevin function.

Now we introduce the following dimensionless variables to simply consider this problem.

$$\bar{r}=r/H, \bar{\psi}=e_0\beta\psi, \bar{\lambda}=\lambda/H, \bar{d}=d/H, \bar{\varepsilon}_r=\varepsilon_r/\varepsilon_p, \lambda=\sqrt{\frac{\varepsilon_0\varepsilon_p k_B T}{2n_b e_0^2}}, \bar{h}=hV_w,$$

$$\bar{V}=V/V_w, \theta=\frac{1}{n_{wb}V_w}, \bar{n}_b=\frac{n_b}{n_{wb}}, p_0\beta|\nabla\psi|=\chi|\nabla\bar{\psi}|, \chi=\frac{p_0}{e_0 H}, |\nabla\bar{\psi}|=\frac{d\bar{\psi}}{d\bar{r}}$$

Using the dimensionless variables, Eq. (5) is transformed into the following equation:

$$\frac{d}{d\bar{r}}\left(\bar{\varepsilon}_r\frac{d\bar{\psi}}{d\bar{r}}\right)=\frac{1}{\bar{\lambda}^2}\frac{\theta}{2}\frac{\exp(\bar{\psi}-\bar{V}_-\bar{h})-\exp(-\bar{\psi}-\bar{V}_+\bar{h})}{\frac{\sinh(\chi|\nabla\bar{\psi}|)}{\chi|\nabla\bar{\psi}|}\exp(-\bar{h})+\bar{n}_b\bar{V}_-\exp(\bar{\psi}-\bar{V}_-\bar{h})+\bar{n}_b\bar{V}_+\exp(-\bar{\psi}-\bar{V}_+\bar{h})}. \quad (7)$$

On the other hand, we consider that an external axial electric field applied to the nanochannel triggers an electroosmotic flow. We assume that the flow is steady, one dimensional, and fully developed. Such a flow field can be caused under the condition that we not only consider a steady electric field but also assume that the electric double layer electrostatics and the ion distribution are not affected by the electroosmotic flow field.

We'll derive a new formula of electroosmotic velocity according to

$$\eta\frac{d^2u}{dr^2}+e_0(n_+-n_-)E_x=0, \quad (8)$$

where $\eta$ and $E_x$ are the dynamic viscosity of water and the employed constant axial electric field, respectively (Sinha et al. 2015).

Using Eq. (3) to replace $n_\pm$, we can finally express Eq. (5) in dimensionless form as

$$\frac{d^2\bar{u}}{d\bar{r}^2} - \bar{E}_x \frac{\theta}{\lambda^2} \frac{\exp(\bar{\psi} - \bar{V}_-\bar{h}) - \exp(-\bar{\psi} - \bar{V}_+\bar{h})}{\frac{\sinh(\chi|\nabla\bar{\psi}|)}{\chi|\nabla\bar{\psi}|}\exp(-\bar{h}) + \bar{n}_b\bar{V}_-\exp(\bar{\psi} - \bar{V}_-\bar{h}) + \bar{n}_b\bar{V}_+\exp(-\bar{\psi} - \bar{V}_+\bar{h})} = 0. \quad (9)$$

with $\bar{E}_x = E_x/E_0$ and $\bar{u} = u/u_0$, where $E_0 = k_B T/(e_0 h)$ and $u_0 = (k_B T/e_0)\varepsilon_0\varepsilon_r E_0/\eta$ respectively mean the electric field due to the potential $k_B T/e$ applied over half nanochannel height and the velocity scale considered equal to the Helmholtz-Smoluchowski velocity that results from a system zeta potential of magnitude $k_B T/e$ in the presence of an electric field $E_0$.

Corresponding electroosmotic velocity $u$ will be obtained by numerically solving Eq. (9) in the presence of the conditions:

$$(\bar{u})_{\bar{r}=\pm 1} = 0, \quad \left(\frac{d\bar{u}}{d\bar{r}}\right)_{\bar{r}=0} = 0. \quad (10)$$

The first equation of Eq. (10) means that at the plane ($\bar{r} = \pm 1$) where the zeta potential is defined, there is no slip for the axial flow velocity, as in (Sinha et al. 2015).

Combining Eqs. (8) and (5) provides the following equation:

$$\frac{d^2 u}{dr^2} = \frac{\varepsilon_0\varepsilon_p E_x}{\eta}\frac{d}{dr}\left(\bar{\varepsilon}_r \frac{d\psi}{dr}\right). \quad (11)$$

Integrating Eq. (11) with respect to $\bar{r}$ and expressing the consequent equation in dimensionless form, we can obtain the following equation:

$$\frac{d\bar{u}}{d\bar{r}} - \bar{\varepsilon}_r \bar{E}_x \frac{d\bar{\psi}}{d\bar{r}} = const. \quad (12)$$

Using the second boundary condition of Eq. (10), we can know that the constant of Eq. (12) should be equal to 0.

Integrating Eq. (12) with const=0 produces the following equation:

$$\bar{u} = \int_0^{\bar{r}} \bar{\varepsilon}_r \bar{E}_x \frac{d\bar{\psi}}{d\bar{r}} dr + const. \quad (13)$$

Using the first boundary condition of Eq. (10) yields the following equation:

$$\bar{u} = \int_0^{\bar{r}} \bar{\varepsilon}_r \bar{E}_x \frac{d\bar{\psi}}{d\bar{r}} d\bar{r} - \int_0^1 \bar{\varepsilon}_r \bar{E}_x \frac{d\bar{\psi}}{d\bar{r}} d\bar{r} = \int_1^{\bar{r}} \bar{\varepsilon}_r \bar{E}_x \frac{d\bar{\psi}}{d\bar{r}} d\bar{r} = \int_{\bar{\psi}_\zeta}^{\bar{\psi}(\bar{r})} \bar{\varepsilon}_r \bar{E}_x d\bar{\psi}. \quad (14)$$

where $\bar{\varepsilon}_r$ is less than unity owing to its definition and $\bar{\psi}_\xi$ is the dimensionless zeta potential at the charged wall ($\bar{r} = \pm 1$) of the nanochannel.

When we neglect the solvent polarization, the relationship between the dimensionless electroosmotic velocity and the dimensionless electrostatic potential is reduced to the corresponding one as in (Chakraborty and Das 2008)

$$\bar{u} = -\bar{\psi}_\xi \bar{E}_x \left(1 - \frac{\bar{\psi}}{\bar{\psi}_\xi}\right) \quad (15)$$

Eq. (15) indicates that when we disregard solvent polarization, the electroosmotic velocity is linearly varied with the electrostatic potential and the magnitude of the electroosmotic velocity is proportional to the difference between electrostatic potential and zeta potential.

Although Eq. (15) is correct only for the case without solvent polarization, we assume that for the case where solvent polarization is considered, the above relation is approximately satisfied by using an appropriate effective permittivity.

2.2. Streaming Potential and Electroviscous effect

We consider the pressure-driven transport of an electrolyte solution in the same charged nanochannel as in section 2.1. A pressure gradient applied along the longitudinal ($x$) direction of the channel causes a fluid flow along the same direction.

For the case, calculation of electrostatic potential is identical with corresponding one of electroosmotic transport, shown in subsection 2.1.

We assume that the flow field consists of the pure pressure-driven transport and the electroosmotic transport induced by the streaming potential $E_s$.

Therefore the governing equation for the velocity field $u_{st}$ can be expressed as

$$\eta \frac{d^2 u_{st}}{dr^2} - \frac{dp}{dx} + e(n_+ - n_-)E_s = 0, \quad (16)$$

where $\frac{dp}{dx}$ is the pressure gradient.

We express Eq. (16) in dimensionless form as follows.

$$\frac{d^2 \bar{u}_{st}}{d\bar{r}^2} - 1 - \frac{1}{2\bar{\lambda}^2} \bar{E}_s (\bar{n}_- - \bar{n}_-) = 0 \quad (17)$$

In Eq. (17), $\bar{u}_{st} = \dfrac{u_{st}}{u_{p,0}}$ (where $u_{p,0} = \dfrac{h^2}{\eta}\dfrac{dp}{dx}$ is pressure-driven velocity scale), $u_r = \dfrac{u_0}{u_{p,0}}$, $\bar{E}_s = \dfrac{E_s}{E_0}$.

The velocity field is solved in presence of the following boundary condition.

$$(\bar{u}_{st})_{\bar{r}=1} = 0;\ \left(\dfrac{d\bar{u}_{st}}{d\bar{r}}\right)_{\bar{r}=0} = 0; \quad (18)$$

Solving $\bar{u}_{st}$ requires the value of the $\bar{E}_s$ obtained by satisfying the condition that the total ionic current per unit length $i$ is zero, i.e.

$$i = 2e\int_0^h (u_+ n_+ - u_- n_-)dr = 0, \quad (19)$$

where $u_\pm$ are the ion migration velocities, expressed as:

$$u_\pm = u_{st} \pm \dfrac{eE_s}{f_\pm} \quad (20)$$

Here $f_\pm$ is the ionic friction coefficient for ions in electrolyte.

As a consequence the dimensionless streaming potential should be obtained by the following equation

$$\bar{E}_s = \dfrac{1}{u_r}\dfrac{\int_0^1 \bar{u}_{st}(\bar{n}_- - \bar{n}_+)d\bar{r}}{R_+\int_0^1 \bar{n}_+ d\bar{r} + R_-\int_0^1 \bar{n}_- d\bar{r}}. \quad (21)$$

We substitute this formula in Eq. (17), and we get an integro - differential equation in terms of $\bar{u}$:

$$\dfrac{d^2\bar{u}_{st}}{d\bar{r}^2} - 1 - \dfrac{1}{2\bar{\lambda}^2}\dfrac{\int_0^1 \bar{u}_{st}(\bar{n}_- - \bar{n}_+)d\bar{r}}{R_+\int_0^1 \bar{n}_+ d\bar{r} + R_-\int_0^1 \bar{n}_- d\bar{r}}(\bar{n}_- - \bar{n}_+) = 0 \quad (22)$$

Eq. (22) is solved numerically in presence of the boundary conditions expressed in Eq. (18). This numerical procedure requires application of suitable iteration procedure; the starting guess profile of the iteration is typically the $\bar{u}_{st}$ profile obtained for the analytical case in our previous study. Once $\bar{u}_{st}$ has been obtained by solving Eq. (22) in presence of the conditions expressed in Eq. (18), we can use Eq. (21) to obtain $\bar{E}_s$.

It is well-known that the streaming potential-induced electroosmotic transport resists the pressure-driven transport. As a consequence the net flow rate is smaller than that of the pure pressure-driven flow rate.

This lowering is known as the electroviscous effect described in terms of an enhanced viscosity ($\eta_{eff}$)

$$\frac{\eta_{eff}}{\eta} = \frac{\int_0^1 \bar{u}_p d\bar{r}}{\int_0^1 \bar{u}_{st} d\bar{r}} \qquad (23)$$

where the dimensionless pure pressure-driven velocity field is $u_p = -\frac{dp}{dx}\frac{h^2}{2\eta}\left(1-\frac{r^2}{h^2}\right)$.

The efficiency of the electrochemomechanical energy conversion is obtained by the following equation , as in (Chanda et al. 2014)

$$\xi = \frac{3}{4}\frac{\bar{E}_s^2 u_r^2 R}{\bar{\lambda}^2} \qquad (24)$$

**3. Results and Discussion**

We denote our approach, which accounts for solvent polarization and non-uniform size effect, as Langevin - Modified - Poisson- Boltzmann (LMPB) approach.

For clarity, we use $T$=300K for all the calculations we study here.

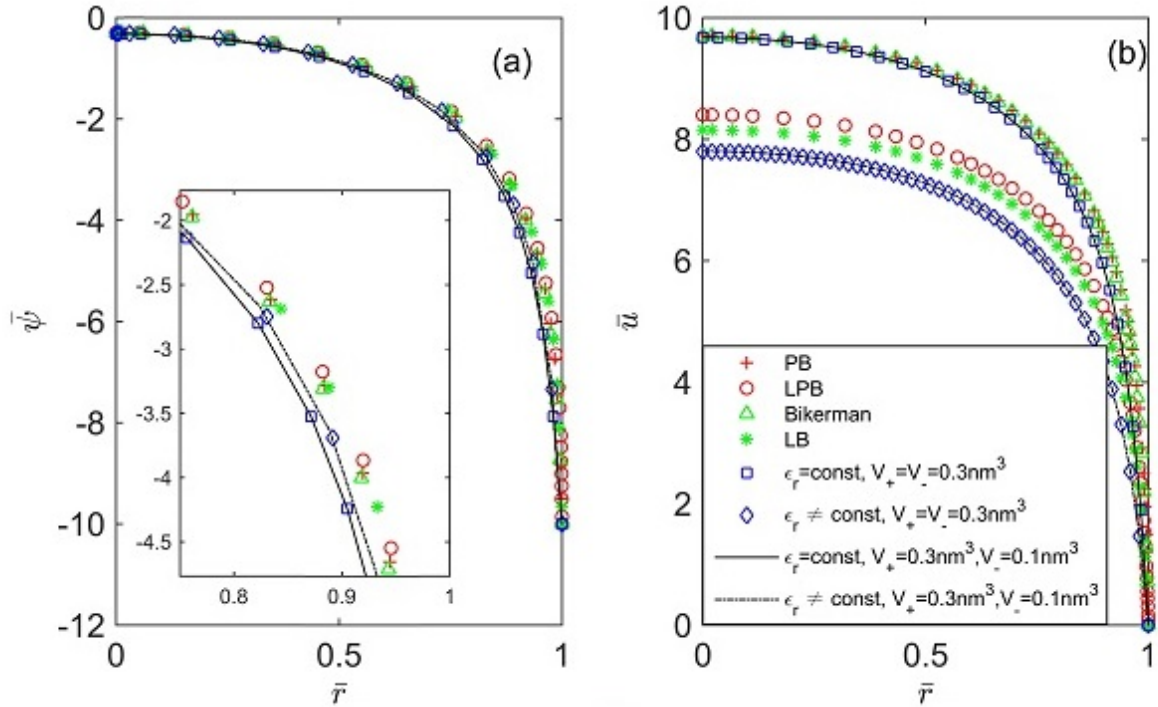

Fig. 2. (Color online) Transverse variation in the dimensionless (a) electrostatic potential and (b) the dimensionless electroosmotic transport calculated by different electric double layer models. Crosses, Circles, Triangles and Squares mean Poisson-Boltzmann(PB), Langevin - Poisson - Boltzmann(LPB), Bikerman and Langevin - Bikerman(LB) models, respectively. A, B, C and D represent MPB ($V_- = 0.1505 nm^3$, $V_+ = 0.1505 nm^3$), LMPB($V_- = 0.1505 nm^3$, $V_+ = 0.1505 nm^3$), MPB ($V_- = 0.1505 nm^3$, $V_+ = 0.230 nm$), LMPB ($V_- = 0.1505 nm^3$, $V_+ = 0.230 nm^3$), respectively. Here $H$=10$nm$, $\overline{\psi}_\xi$ =10 and cb=0.01M.

Fig. 2(a) shows the transverse variation in electrostatic potential across the entire nanochannel with height of $H = 10$ $nm$ for different electric double layer models. The cases for point-like charges are represented by crosses (PB) and circles (Langevin-Poisson-Boltzmann (LPB)). For the cases where water molecules and ions have the same size, triangles and asterisks represent Bikerman and LB. In addition, the cases for ($V_+ = V_- = 0.1505$ $nm^3$, $V_w = 0.03$ $nm^3$) are illustrated by squares and diamonds. The cases for ($V_- = 0.1505 nm^3$, $V_+ = 0.230$ $nm^3$, $V_w = 0.03$ $nm^3$) are represented by solid line and solid-cross line, respectively. Here $V_- = V_+ = 0.1505$ $nm^3$ corresponds to $Cl^-$, $Br^-$, $I^-$, $Rb^+$, $K^+$ and $Cs^+$ having hydrated radius 3.3 Å. $V_+ = 0.230$ $nm^3$ means the volume of $Li^+$ having hydrated radius 3.8 Å. We refer to the classical experimental work of (Nightingale 1959).

The first finding in Fig. 2(a) is that at a given zeta potential, solvent polarization invariably provides a lower electrostatic potential across the entire nanochannel as compared to the case without solvent polarization. This results from the fact that orientational ordering of water dipoles lowers the permittivity of electrolyte solution. In (Das et al 2012a), the authors obtained such results for the case of electrolytes containing point-like ions. The other finding of Fig. 2(a) is that enhancement of ion size invariably enhances the magnitude of the electrostatic potential. This is understood by noting that a large counterion size possesses weak screening property and thereby gets high electrostatic potential compared to the case of smaller sizes of counterions. As a consequence, the electroosmotic velocity for the case of a large counterion size is smaller than for the case of a smaller counterion size. In fact, as indicated in (Das and Chakraborty 2011), an enhancement in counterion size induces thickening of electric double layer and results in the extension of effective electric double layer overlap and a significantly enhanced value of the channel centerline potential.

Fig. 2(b) shows the electroosmotic velocity profile as a function of dimensionless transverse location. It is seen that an increase in counterion size invariably causes a decrease in electroosmotic velocity across the entire nanochannel, irrespective of accounting for solvent polarization. Let's consider the case of constant permittivity. In fact, although for the case of a large counterion size, the magnitude of electrostatic potential is larger than for a smaller counterion size, the difference between the zeta potential and the electrostatic potential becomes smaller than for the latter case, owing to the condition of a given zeta potential. On the other hand, Eq. (15) indicates that the electroosmotic velocity is proportional to the difference between zeta potential and the electrostatic potential.

We can know that for the case with solvent polarization, the linear coefficient is lower than for the case without solvent polarization, thereby lowering the permittivity near the wall of the nanochannel. Fig. 2(b) demonstrates that such a trend yields not only smaller electroosmotic velocities but also a larger difference in electroosmotic velocity between the cases of different counterion sizes.

In a word, it is illustrated that solvent polarization amplifies ion size effect.

Fig. 3(a)-(d) are the transverse variations in counterion number density, coion number density, water molecule number density and dielectric permittivity, respectively. All the parameters are the same in Fig. 2(a) and (b).

Fig. 3(a) shows that a larger counterion size produces a weaker accumulation of counterions near a wall of the nanochannel owing to steric effects. On the other hand, solvent polarization lowers the counterion number density owing to lowering of electrostatic potential, as explained in Fig. 2(a). In a similar way, we can anticipate that solvent polarization allows the coion number density to be higher than for the case without solvent polarization.

Fig. 3(b) displays that for both the cases where counterion sizes have the identical size and coion sizes are different, coion number density for the case with coions of the smaller size is higher than for the case with coions of the larger size.

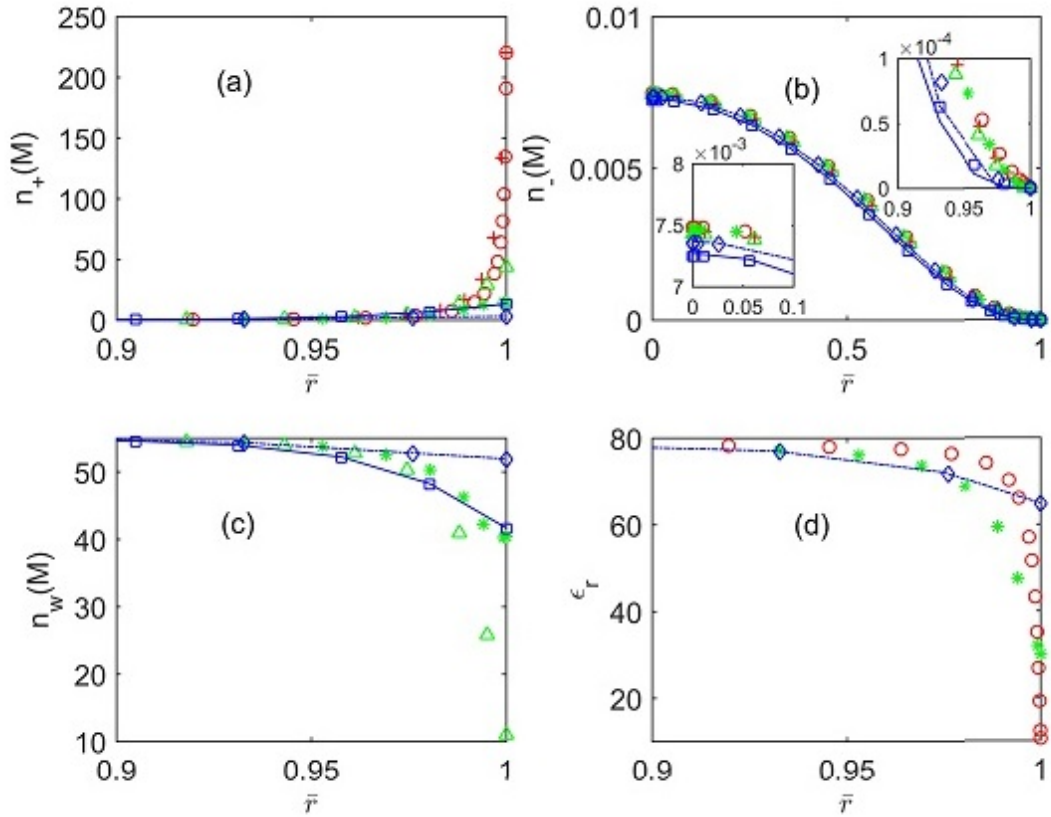

Fig. 3. (Color online) Transverse variation in the dimensionless (a) counterion number density, (b) coion number density and (c) water dipole number density and (d) the relative permittivity for different electric double layer models. All the parameters are the same as in Fig. 2.

Fig. 3(c) shows that water molecule number density for counterions of a large size is higher than for smaller size of counterions owing to the volume conservation condition. In the same way, we can prove that solvent polarization enhances water molecule number density. Fig. 3(d) shows that dielectric permittivity has the same trend as in water molecule number density due to Eq. (6).

Fig. 4(a) and (b) depict the electrostatic potential profile and electroosmotic velocity profile for different bulk ion number densities and zeta potentials calculated by using LB and LMPB approaches.

Fig. 4(a) demonstrates that the lower bulk ion number density produces the higher electrostatic potential across the entire nanochannel than that of the higher bulk ion number density. This is attributed to the fact that a low ion number density requires a high electric force to screen an electric charge of a wall of nanochannel compared to the case of a higher bulk ion number density.

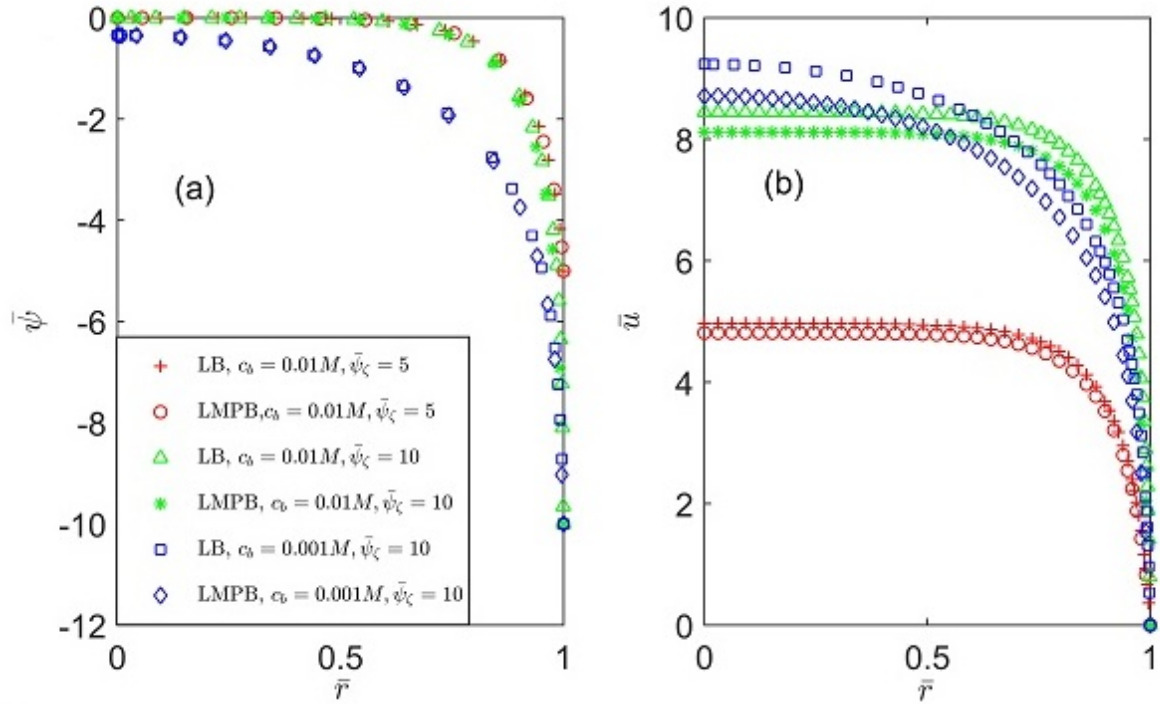

Fig. 4. (Color online) Transverse variation in the dimensionless (a) electrostatic potential and (b) the dimensionless electroosmotic velocity calculated by LB and our approach (LMPB) for different combinations of the bulk ion number densities and the dimensionless zeta potentials. Here $H=30nm$ and $V_+=V_-=0.3nm^3$.

Fig. 4(b) shows that the electroosmotic velocity approaches a saturated value in the vicinity of the centerline. Let's consider in more detail. For both the cases where solvent polarization is not considered and the electrostatic potential at the centerline of nanochannel is nearly zero, the ratio of the magnitudes of the electroosmotic velocity at the centerline should be determined by Eq. (15).

However, we verify that for the case where solvent polarization is considered, a high zeta potential gives a low dielectric permittivity of the electrolyte solution near the charged surface owing to orientational ordering of water dipoles and thereby ensures a low electroosmotic velocity compared to the case without solvent polarization. As a consequence, the ratio of the electroosmotic velocities $\bar{u}_{\bar{\psi}_\varsigma=10}/\bar{u}_{\bar{\psi}_\varsigma=5}$ is lower than the ratio of the zeta potentials 10/5.

Fig. 4(b) also shows that for a given zeta potential, the electroosmotic velocity obtained by using LB approach is higher than corresponding one by using LMPB. This is explained by the fact that as we will see in Fig. 5(d), when ion size increases, a decrease in permittivity of electrolyte solution due to solvent polarization is more pronounced due to increasing of ion occupying volume and thereby ensures $\bar{u}_{LMPB} < \bar{u}_{LB}$.

Fig. 4(b) also shows that an increase in zeta potential enhances the difference in the magnitude of the electroosmotic velocities obtained by using LB and LMPB. This is attributed to the fact that as we will see in Fig. 5(d), when zeta potential increases, a decrease in permittivity of electrolyte solution due to solvent polarization is enhanced and consequently the difference in the magnitude of the velocity between the cases increases. Fig. 4(b) also demonstrates that for a given zeta potential, a high bulk ion number density provides a large magnitude of electroosmotic velocities at the centerline compared to the case with a smaller density. This is attributed to the fact that for low bulk ion number density, although the difference between zeta potential and the electric potential at the centerline gets smaller than for a high bulk ion number density, a decrease in permittivity of electrolyte solution is weaker than that for high bulk ion number density.

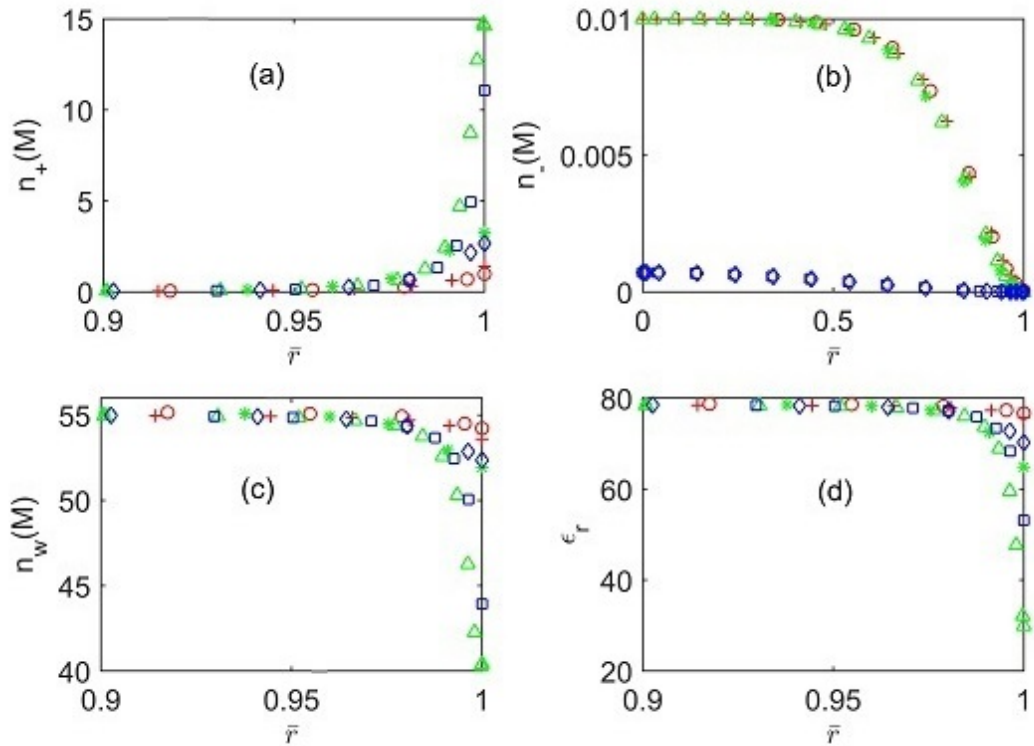

Fig. 5. (Color online) Transverse variation in the dimensionless (a) counterion number density, (b) coion number density and (c) water dipole number density and (d) the relative permittivity calculated by LB and our approach for different bulk ion number densities and zeta potentials. All the parameters are the same in Fig. 4.

Fig. 5(a)-(d) depict transverse variations in counterion number density, coion number density, water molecule number density and permittivity for different bulk ion number densities and zeta

potentials obtained by using LB and LMPB approaches.

In Fig. 5(a), we can know that the counterion number density for the case of a low bulk ion number density is lower than corresponding one for the case of a higher bulk ion number density.

From the above fact, it is proved that as seen in Fig. 5(c), the water molecule number density for the former case is lower than for the latter case, thereby implying the same trend in the relative permittivity as in water molecule number density Fig. 5(d).

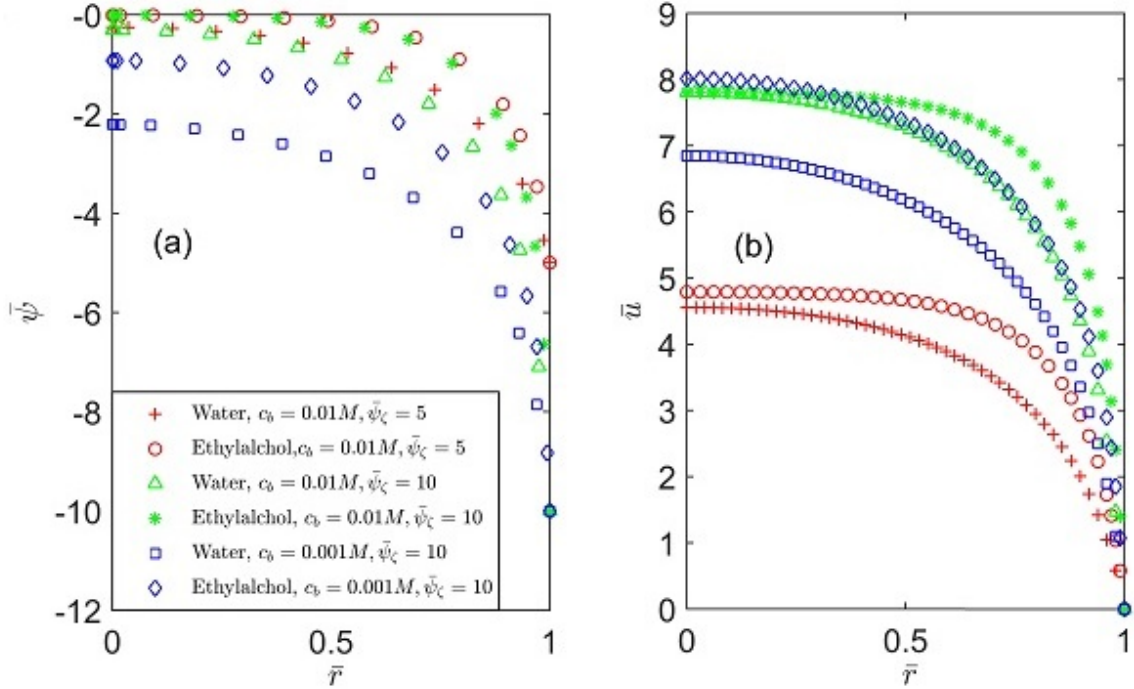

Fig. 6. (Color online) For both the cases where water or ethylalcohol is solvent, transverse variation in the dimensionless (a) electrostatic potential and (b) the dimensionless electroosmotic velocity calculated by our approach for different bulk ion number densities and zeta potentials. Here $H$=20$nm$ and $V_+$=$V_-$=0.3$nm^3$.

Fig. 6(a) and (b) show the electrostatic potential profile and electroosmotic velocity profile for different bulk ion number densities and zeta potentials calculated for electrolytes where solvent is water or ethylalcohol. We use the volume of an ethylalcohol molecule of 0.0968$nm^3$, since for pure ethylalcohol solution the molar weight of ethylalcohol is 46g/mol and the mass density is 786kg/$m^3$. In fact, (Chang et al 2010) studied electroosmosis in microchannel with low permittivity liquids like ethanol or butanol.

Fig. 6(a) shows that for a low bulk ion number density, the difference in electrostatic potential between the electrolytes having water or ethylalcohol as solvent is large compared to the case of a high bulk ion number density.

As can be shown in Fig. 6(b), such a fact is also satisfied in the electroosmotic velocity profile owing to Eq. (14).

On the other hand, Fig. 6(b) shows that increasing the zeta potential induces a decrease in the difference between the electroosmotic velocities at the centerline for the two electrolytes containing water or ethylalcohol as solvent. This is understood by the fact that the difference between the electrostatic potentials at the centerline for the two cases plays no role in determining the difference between the electroosmotic velocities since the variation in permittivity for the two cases cancel out with the effect.

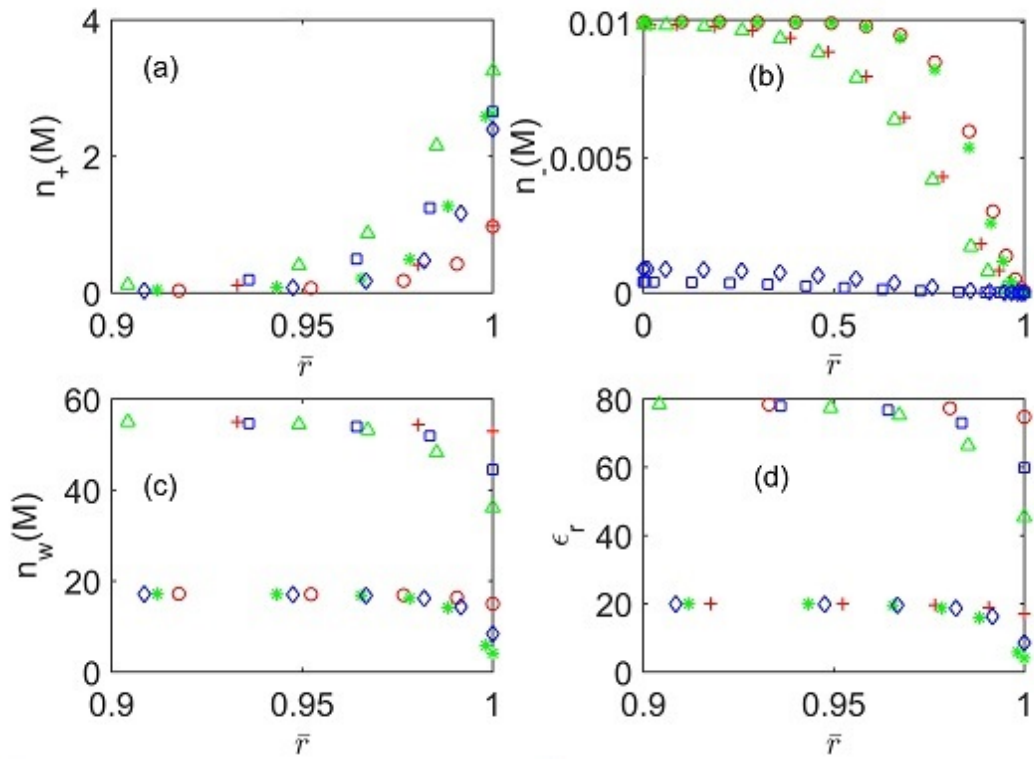

Fig. 7. (Color online) For both the cases where water or ethylalcohol is solvent, transverse variation in the dimensionless (a) counterion number density, (b) coion number density and (c) water dipole number density and (d) the relative permittivity calculated by our approach for different bulk ion number densities and zeta potentials. All the parameters are the same in Fig. 6(a) and (b).

Fig. 7(a)-(d) depict the transverse variation in dimensionless counterion number density, coion number density, water molecule number density and the relative permittivity for different combinations of bulk ion number densities and zeta potentials calculated for electrolytes with water or ethylalcohol.

Fig. 7(a) illustrates that the counterion number density for the electrolyte with water is larger than that for the electrolyte with ethylalcohol due to the same reason for electrostatic potential.

Fig. 7(b) displays that the coion number density for the former case is lower than that for the latter case. Fig. 7(d) exhibits that the permittivity variation for ethylalcohol electrolyte is higher than that for the case for the aqueous electrolyte, thereby implying the same trend in solvent number density Fig. 7(c).

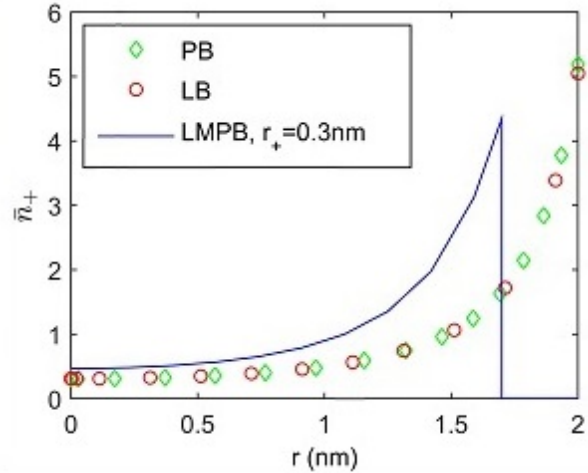

Fig. 8. (Color online) Transverse variation in the dimensionless sodium ion number density. Here the nanochannel half height is $2nm$ and the surface charge density is $\sigma=-0.095C/m^2$. We use the bulk sodium ion concentration of $0.55M$, which is equal to that for the electrolyte containing 20 sodium ions and 2067 water molecules.

Fig. 8 compares the results of the present theory, PB and LB for the case corresponding to the molecular dynamics simulation determining the counterion spatial distribution in (Kim and Darve 2006). Here the simulation is governed by the following parameters: the nanochannel half height is $2nm$ and the surface charge density is $\sigma=-0.095C/m^2$. The electrolyte contains 20 sodium ions and 2067 water molecules, while the concentration of the negatively charged ions is considered as zero and T=300K. In particular, it is important to note that we slightly modify the present approach, including the closest approach of the counterion radius to the wall of the nanochannel. We find that our model with hydrated sodium ion radius of $0.358nm$ much better reproduces the molecular dynamics simulation result than PB or LB. Although there exists a slight discrepancy between our model and MD simulation (Kim and Darve 2006), such a behavior is mainly due to the roughness of nanochannel wall used in the simulation. As a consequence, we anticipate that the simultaneous consideration of non-uniform ion size effect and solvent polarization ensures a much more realistic description of the electric double layer electrostatics and electroosmotic transport.

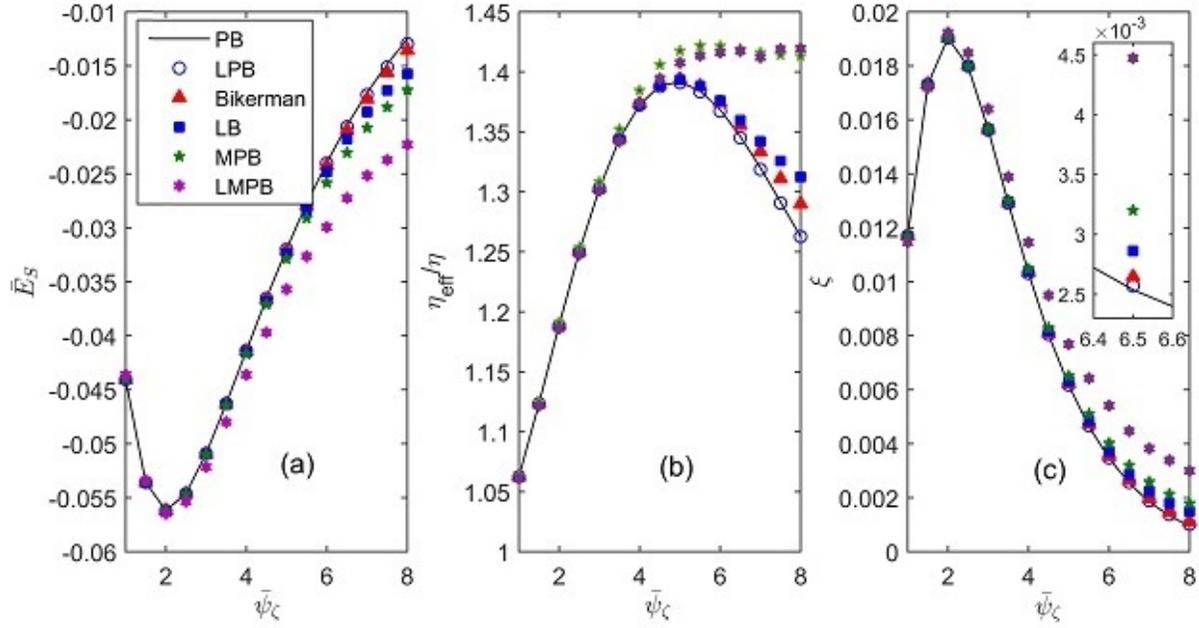

Fig. 9 (Color Online) Variation of (a) the dimensionless streaming potential, (b) the dimensionless effective viscosity ($\eta_{eff}/\eta$), quantifying the electroviscous effect and (c) electrochemomechanical efficiencies with dimensionless zeta potential for different approaches (PB, LPB, Bikerman, LB, MPB and LMPB). For MPB and LMPB approaches, we use V$^-$ = 0.1505nm³, V$^+$ = 0.230 nm³ for ionic volumes. $u_r = 1, R_+ = R_- = 3$, H=5nm.

Fig. 9(a) shows the variation of the numerically calculated streaming potential with different approaches for nanochannels.

For low zeta potentials, all the approaches provide the same results for streaming potential.

Firstly, one can discover that an increase in counterion volume results in an enhancement in magnitude of streaming potential, for all the cases whether or not considering solvent polarization.

This is explained by the two following facts. First, as mentioned above, steric effect of ions yields an extension of effective electric double layer thickness. Second, Eq. (21) means that streaming potential for the case with a thin electric double layer thickness is smaller than that of a thicker electric double layer, since the dimensionless velocity field vanishes near a nanochannel wall and increases towards the centerline of the channel.

The second issue is that solvent polarization invariably increases the magnitude of streaming potential. In fact, solvent polarization allows counterion number density to slowly vary. As a result, the magnitude of $\bar{u}_{st}(\bar{n}_- - \bar{n}_+)$ is larger than that of the case when not considering solvent polarization, thereby implying increase in magnitude of streaming potential.

Fig. 9(b) shows the variation of the dimensionless effective viscosity with different approaches for nanochannels. First, it is shown that an increase in ion size yields an increase in effective viscosity due to the decrease of corresponding flow. Second, it should be strongly noted that for a low zeta potential solvent polarization effect increases effective viscosity while for a high zeta potential, it decreases the viscosity. This is attributed to combined effect of the increase in magnitude of streaming potential and changed spatial profile of $(\bar{n}_- - \bar{n}_+)$ when we consider solvent polarization.

Fig. 9(c) shows the variation of electrochemomechanical efficiencies with different approaches for nanochannels. From Eq. (27), for a given $\bar{\lambda}$ electrochemomechanical efficiency has similar behaviors to ones of streaming potential.

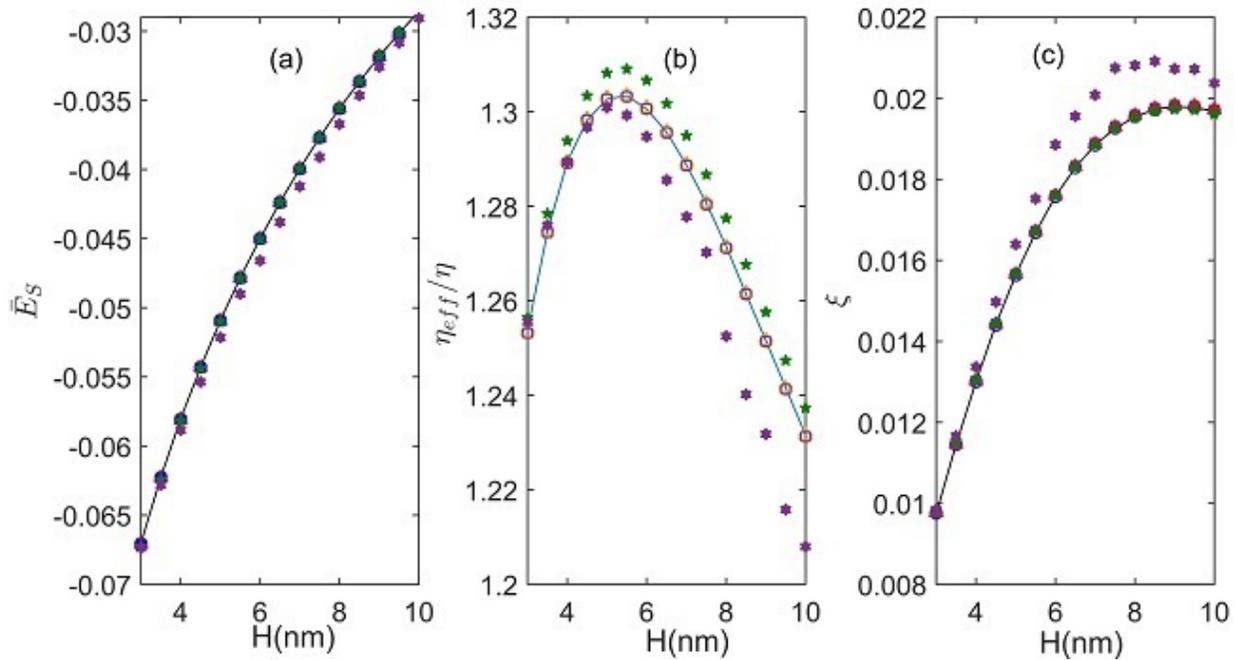

Fig. 10 (Color Online) Variation of (a) the dimensionless streaming potential, (b) the dimensionless effective viscosity ($\eta_{eff}/\eta$), quantifying the electroviscous effect and (c) electrochemomechanical efficiencies with nanochannel half height (H) for different approaches (PB, LPB, Bikerman, LB, MPB and LMPB). For MPB and LMPB approaches, we use $V^- = 0.1505$ nm$^3$, $V^+ = 0.230$ nm$^3$ for ionic volumes. $u_r = 1$, $R_+ = R_- = 3$ and $\bar{\psi} = 3$.

Fig. 10(a), (b) and (c) show the variation of the numerically calculated streaming potential,

effective viscosity and electrochemomechanical efficiency with nanochannel half height for different approaches for nanochannels, respectively.

From all the figures, we can confirm that our findings from Fig. 8(a), (b) and (c) validate in the total range of nanochannel height.

In particular, Fig. 10(b) displays that for a low zeta potential, solvent polarization effect (LMPB) is enhanced with counterion size and thus results in a lower viscosity than that of other approaches (PB, LPB, Bikerman and LB) with a smaller counterion sizes.

**4.** Conclusions

We have provided to study solvent polarization effect and non-uniform size effect in electroosmotic transport in a nanochannel. It has been shown that in the presence of a given zeta potential, solvent polarization effect has been found to lower the electrostatic potential and thereby ensures the lowering of electroosmotic velocity. Also, we have demonstrated that solvent polarization effect on electroosmotic transport is more enhanced with increasing zeta potential. On the other hand, in the presence of a given zeta potential, an increase in counterion size enhances the electrostatic potential across the nanochannel, which in turn augments the electroosmotic transport for the cases considering or not solvent polarization. In addition, a lower bulk ion number density causes not only an increase in permittivity near a wall of the nanochannel but also an increase in electrostatic potential across the nanochannel compared to the case of a high bulk ion number density. We have found that the difference in electroosmotic velocity between the cases with different counterion sizes at the centerline for a low bulk ion number density becomes smaller than the corresponding one for the case of a higher bulk ion number density. Moreover, in the presence of low bulk ion concentration, the electrostatic and electroosmotic properties of aqueous electrolytes strongly differ from those of electrolytes containing ethylalcohol as compared to the case of a higher bulk ion concentration. We have concluded that solvent polarization and ionic size should be explicitly considered when we study streaming potential and electroviscous effect.

This study not only unravels a basically important connection between electric double layer electrostatics and electrokinetic transport by simultaneously considering solvent polarization effect and ion size effect, but also promises a new mechanism of electroosmotic flow control that will potentially affect the applications involving liquid and species transport.